\documentclass[fleqn,usenatbib]{mnras}

\usepackage[T1]{fontenc}

\DeclareRobustCommand{\VAN}[3]{#2}
\let\VANthebibliography\thebibliography
\def\thebibliography{\DeclareRobustCommand{\VAN}[3]{##3}\VANthebibliography}

\usepackage{graphicx}	
\usepackage{amsmath}	
\usepackage{amssymb}	
\usepackage{lineno}

\usepackage{soul}
\usepackage{newtxtext,newtxmath}

\title[Exp. discs in a Markov chain]{ 
Exponential 
 galaxy discs as the quasi-stationary distribution in a Markov chain model simulating stellar scattering
 }

\author[J. Wu et al.]{
Jian Wu,$^{1}$\thanks{E-mail: jianwu@alumni.iastate.edu (JW), curt@iastate.edu (CS)}
Curtis Struck,$^{1}$\footnotemark[1]
Bruce G. Elmegreen$^{2}$
and Elena D'Onghia$^{3}$
\\
$^{1}$Department of Physics and Astronomy, Iowa State University, 2323 Osborn Dr., Ames, IA 50011, USA\\
$^{2}$IBM Research Division, T.J. Watson Research Centre, 1101 Kitchawan Road, Yorktown Heights, NY 10598, USA\\
$^{3}$Department of Astronomy, University of Wisconsin-Madison, 475 N Charter St, Madison, WI 53706, USA
}

\date{Accepted 2023 April 19. Received 2023 March 21; in original form 2022 November 6}

\pubyear{2023}

\begin{document}
\label{firstpage}
\pagerange{\pageref{firstpage}--\pageref{lastpage}}
\maketitle

\begin{abstract}
Previous models have shown that stochastic scattering of stars in a two-dimensional galaxy disc 
can generate a  time-independent  surface density distribution that is an exponential divided by radius when a constant inward scattering bias 
is present. Here we show, using a Markov chain model,  that similar profiles result from an outward scattering bias, although the disc surface density decreases slowly with time because of a net stellar outflow.  The trend towards a near-exponential surface profile is robust, as it exists even if the scattering intensity has moderate radial and time dependences, subject to some limitations on the scattering rates discussed in the text. The exponential scale length of the pseudo-equilibrium disc depends on the scattering bias, the scattering length, and the size of the disc where scattering  is important.

\end{abstract}

\begin{keywords}
galaxies: disc -- galaxies: evolution -- galaxies: kinematics and dynamics
\end{keywords}



\section{Introduction}

The exponential shape of a stellar disc is common in many types of galaxies,
such as spirals,  dwarf irregulars, and dwarf ellipticals  \citep{boroson1981,ichikawa1986,Andredakis1995MNRAS.275..874A,Patterson1996,deJong1996A&A...313...45D,Simard2011ApJS..196...11S,Herrmann2013AJ....146..104H}. It is observed  at both low  and  
high redshifts \citep{Law2012ApJ...745...85L,Patricio2016MNRAS.456.4191P,Bowler2017}. Even clumpy galaxies with multiple overdense regions in their disc hold an exponential stellar profile 
\citep{Elmegreen2005ApJ...634..101E,Bournaud2007,Shibuya2016ApJ...821...72S}.  In spite of the ubiquity of 
exponential discs, their formation is not well understood.

The emergence of an exponential disc can come from  initial and continuing conditions related to cosmic accretion combined with internal processes that shift stars
radially in a galaxy. These processes can occur through  various mechanisms.
Spiral arms produce stellar radial migration near the corotation radius \citep{Sellwood2002,Roskar2008ApJ...675L..65R, Vera-Ciro2014ApJ,Daniel2018MNRAS.476.1561D}.  
Massive clumps in the disc, such as giant 
molecular clouds, can move stars radially by their gravity when they encounter
stars \citep{Elmegreen2013ApJ...775L..35E,Struck2017, 2020Wu}. If a galaxy has a central bar, the  gravitational force from
the bar can modify stellar orbits outside the bar and move the stars radially \citep{hohl1971, Debattista2006ApJ...645..209D}. When two galaxies merge, redistribution of stars is inevitable due to potential changes and tidal forces   \citep{Pe2006ApJ...650L..33P}.

Since stars are born from gas clouds, processes that relocate gas in the disc
also give rise to changes in the stellar radial profile. Regarding gas rearrangement, gas can be blown out of the disc by stellar feedback from
massive stars and then fall back to the disc \citep{Normandeau1996Natur.380..687N,Pavel2012ApJ...760..150P,Li2013RAA....13..921L}. \cite{Struck2018ApJ...868L..15S} showed that such fountains can produce an exponential radial profile. Gas also moves radially  into an exponential  when it transfers angular momentum due to  viscosity  proportional to the star formation rate \citep{Lin1987,Wang2022ApJ...927..217W}.  Furthermore, models show that cosmic gas can 
flow into the halo of a galaxy and gradually condense onto the galactic disc  \citep{ Hobbs2013MNRAS.434.1849H,Bustard2016},
making changes to the radial distribution of gas.  These changes can preserve an initial exponential under reasonable conditions, although the scale length may shrink if the accreted gas has lower specific angular momentum than the disc at the radius where it accretes \citep{Elmegreen2014ApJ...796..110E}.

To build up a connection between an exponential disc and  direct radial shifts of stars or indirect shifts via gas rearrangement, \citet{Elmegreen2016ApJ...830..115E} used  a stochastic model which assumes
a fixed radial bias in stellar scattering. They
showed that
scattering with a slight inward bias makes an  equilibrium profile
that is  an exponential divided by radius, $e^{-r/h}/r$, which, for galaxies, would be
indistinguishable from an exponential beyond one scale length, $h$.  Inner disk profiles are usually observed to be steeper within the first scale length \citep{Bouquin2018ApJS..234...18B}, which is the inner one-magnitude, but it is difficult to disentangle the possible effects of a disk upturn from a bulge or pseudobulge as the combined colors typically get redder there too \citep{bakos08}.
This mathematical form is fundamental for a disc geometry, like the Gaussian form is fundamental for one-dimensional scattering in a Galton board. The basic requirement for the exponential is biased scattering toward a reflecting wall; such a wall is automatic in circular coordinates because radius cannot be negative. After each particle experiences a sufficient number of scatterings,
the exponential scale length derived in the model is $0.5\lambda/(q-p) $  , where
$\lambda$ is the  scattering mean free path, and $p$ and $q$ are the 
outwards and inwards scattering probabilities respectively.
Their model requires that $p$ + $q$ =1, so particles cannot stay still during one scattering time.
They also showed that double exponentials with Type II or Type III profiles  (see \citet{Freeman1970ApJ...160..811F} and \citet{Erwin2005ApJ...626L..81E} )
can be generated if the inwards scattering bias assumed in the outer disc is
different from the inner disc. Other work that explains the formation of an exponential includes \citet{herpich2017MNRAS.467.5022H}, \citet{Struck2019MNRAS.489.5919S}, and \citet{  Marr2020Galax...8...12M}, although most of them take a perspective of maximum entropy principle. The maximum entropy approach does not contradict the stochastic scattering model, as the random scattering could be viewed as an instance of how the maximum entropy is realized.

\citet{Wu2022MNRAS.517.4417W} (hereafter Paper I) studied orbits and angular momentum changes of stars 
in simulations where exponential discs form due to close star-clump encounters.
They found that an inwards scattering bias is not statistically significant in many galactic radii, and suggested that an outwards scattering bias is possible if a clump profile with a shallow density gradient is used in the simulations.  Following 
this suggestion, we want to explore ways to produce an exponential disc in cases
where no bias or a slight outwards bias is present during scatterings. 

In this paper, we introduce a Markov chain model, simulating stellar scattering in 
a galactic disc. Basic assumptions and the initial stellar distribution used in the model  are presented in Section~\ref{sec:g model}. Stellar profile evolution with different directions of radial scattering bias is given in Section~\ref{sec:g results}. Factors that affect the profile evolution and insights on how to choose model parameters for a specific scattering mechanism are discussed in Section~\ref{sec:g discussion}. The main results are summarized in Section~\ref{sec:g summary}.

\section{Model}
\label{sec:g model}

The basic model considers an initial disk of stars with a radial profile that deviates significantly from an exponential, such as a flat profile, and evolves it through stellar scattering to a quasi-stationary state. Real galaxies start with more tapered profiles as a result of accretion, mergers, and interactions, and the internal scattering process considered here is likely to adjust this profile by only a small amount. The primary assumption is that the quasi-equilibrium profile for the extreme model is about the same as the quasi-equilibrium profile of a real galaxy, regardless of the different histories.

We use a discrete-time discrete-space Markov chain to study the evolution of  the stellar distribution in  a galactic disc. We divide the disc into concentric annuli, each with a fixed width of $d$.
Assuming a maximum disc size $R$, the number of annuli in which we count stars is $N = R/d$. Let $n_i(t)$ denote the number of stars in the $i$-th annulus at time $t$, where $i$ = 1, 2, 3, ..., $N$, and $i$ = 1 corresponds to the innermost annulus. After each timestep, we assume that a star can stay in its annulus or go to an adjacent one.
For a star in the $i$-th annulus, where $i\neq$ 1 or $N$, the probabilities of moving to the
($i$+1)-th annulus and the
($i$-1)-th annulus after every timestep are $p_i$ and $q_i$, respectively, and the probability of not moving is 1-$p_i$-$q_i$. For a star in the first annulus, it is not 
allowed to move inward. In this case, the probabilities of moving to the
second annulus is $p_1$, and the probability of not moving is 1-$p_1$. This is equivalent to a reflecting inner boundary condition.
For a star in the $N$-th annulus, we define the ($N$+1)-th "annulus"  as the region beyond  the radius  $R$ so that the regular transition rule will work here too. Since $R$ is the maximum radius of the disc, stars in the ($N$+1)-th "annulus" are treated as escaping
from the galaxy. Therefore, stars in the ($N$+1)-th "annulus" stay there forever. Hence the ($N$+1)-th "annulus" is an absorbing state. 


As real galactic discs may keep growing bigger and do not have a physical edge, the model disc edge $R$ should be interpreted as a radius beyond which stars are unable to scatter back. The inability to scatter inwards can be caused by processes happening beyond $R$, such as the number of scattering centers plummeting at $R$.  For example, if gas clouds scatter, $R$ refers to the outer regions that just have smooth gas so they
stop scattering. When we say "stars escaping from the galaxy" or "star loss" later in this paper, these stars may still be in the galaxy, but have moved outside of the model disc edge. 

 Unless otherwise stated, we assume that the probabilities of moving to adjacent annuli are independent of the location of the annulus. Hence we drop the subscript $i$ in $p_i$ and $q_i$. The evolution of
the stellar distribution is described by an $N$+1 by $N$+1 transition matrix $P$ as the following:
\begin{equation}
P =
    \begin{pmatrix}
1-p & p & 0 & 0 &...& 0 & 0 & 0\\
q & 1-p-q & p & 0  & ...& 0 & 0 & 0\\
0 & q & 1-p-q & p  &... & 0& 0 & 0\\
. & . & . & .  &  & .& .& .\\
. & . & . & .  &  & .& .& .\\
. & . & . & .  &  & .& .& .\\
0 & 0 & 0 & 0  &... & q & 1-p-q & p \\
0 & 0 & 0 & 0 &... & 0 & 0 & 1
\end{pmatrix},
	\label{eq:trans matrix}
\end{equation}
where the $i$, $j$ entry is the probability of a star going from the $i$-th annulus
to the $j$-th annulus during one timestep. 
If $\pi(t)$ is a row vector representing the numbers of stars in each annulus at time $t$, we have 
\begin{equation}
\pi(t+1)=\pi(t)P.
	\label{eq: matrix evolution}
\end{equation}
Here, we made an assumption that $p$, $q$, and the matrix $P$ do not
depend on time $t$. 

Unless otherwise stated, the disc radius $R$ and the annulus width $d$ are chosen as 15 kpc and 1 kpc, respectively. If the radial transition of stars is caused by clumps in the disc, the shift of 1 kpc can be achieved if a star meets a clump of $10^6$- $10^7M_{\sun}$ (Paper I).  Thus, based on the N-body results of that paper, we adopt this as the typical scattering length for a massive clump.
With $R$ = 15 kpc and $d$ = 1 kpc, the number of annuli $N$ is 15.   Initially, 100 000 stars are  evenly placed in the innermost 10 annuli and the rest of the annuli are empty. Since the  initial star number is constant in  the inner 10  radial bins,
the stellar surface density is proportional to $1/r$ for $r \leq 10$ kpc. 

Time is measured by the number of timesteps away from the starting time.

To make the Markov chain model generally applicable, we do not tie radial scattering of stars to a certain process at this point. Scattering can be caused by encounters with massive clumps, e.g. resonant scattering at corotation due to spiral arms, or the bar. We will connect stellar scattering to a  
specific mechanism in Section~\ref{sec:g connecting} and explain how the model parameters, such as $p$ and $q$, are related to properties of a galaxy.

In a real galaxy, radial scattering of stars is accompanied by angular momentum transport and changes in stellar velocity dispersion. An arbitrary choice of $p$ and $q$ can make the total stellar angular momentum non-conserved. This may not be a serious issue as unbalanced angular momentum can go to gas or dark matter of the galaxy. Changes in velocity dispersion may cause scattered stars and unaffected stars in the same annulus to have different probabilities of getting scattered during next time step. We can partially solve this problem by treating $p$ and $q$ as a function of time, as discussed in Section~\ref{sec:g time-dep}, while keeping stars undistinguished in an annulus. 


\section{Results}
\label{sec:g results}

In this section, we show the evolution of the stellar distribution with
different choices of the $(p,q)$ pair in the model.
Before that, we present the definition and the property of quasi-stationary
distributions.

\subsection{The quasi-stationary distribution of the evolution}
\label{sec:g quasi distrib}

In our Markov chain model, it is easy to see that all the annular populations,
except the ($N$+1)-th one, are transient states. According to 
the properties of Markov chains, the limiting distribution is that
all the stars are in the absorbing state.  Hence, all the stars
 move outside of model disc edge. 
The time to reach this limiting distribution is enormous. 
Our interest in this model is to study quasi-stationary distributions satisfying the condition that the multi-term ratios
$n_1(t) : n_2(t) : ... : n_N(t)$ are unchanged with time, although
each $n_i(t)$ decreases with time.

From the definition of the row vector $\pi(t)$, we know $n_1(t), n_2(t), ..., n_N(t)$ are the first $N$ entries in $\pi(t)$.
Using Equation \ref{eq: matrix evolution}, it is easy to obtain the following:
\begin{equation}
\begin{cases}
     n_1(t+1) - n_1(t) = -pn_1(t)+qn_2(t), \\
    n_2(t+1) - n_2(t) = pn_1(t)-(p+q)n_2(t)+qn_3(t), \\
    n_3(t+1) - n_3(t) = pn_2(t)-(p+q)n_3(t)+qn_4(t), \\
    ... \\
    n_{N-1}(t+1) - n_{N-1}(t) = pn_{N-2}(t)-(p+q)n_{N-1}(t)+qn_{N}(t),\\
    n_N(t+1) - n_N(t) = pn_{N-1}(t)-(p+q)n_N(t).
   \end{cases}
	\label{eq: derive equilibrium equation}
\end{equation}
When a quasi-stationary distribution is reached at time $t$, we have  
\begin{equation}
n_1(t+1) : n_2(t+1) : ... : n_N(t+1) = n_1(t) : n_2(t) : ... : n_N(t) ,
	\label{eq: derive equilibrium equation2}
\end{equation}
which gives the following relation
\begin{equation}
\frac{n_1(t+1)-n_1(t)}{n_1(t)} = \frac{n_2(t+1)-n_2(t)}{n_2(t)} = ... = \frac{n_N(t+1)-n_N(t)}{n_N(t)} .
	\label{eq: derive equilibrium equation3}
\end{equation}
In this scenario, we let $\beta(t)$ denote the fractions in Eqution \ref{eq: derive equilibrium equation3}, i.e.
\begin{equation}
\beta(t) :=\frac{n_1(t+1)-n_1(t)}{n_1(t)}.
	\label{eq: beta}
\end{equation}
Adding up all the numerators and all the denominators, Equation \ref{eq: derive equilibrium equation3} gives
\begin{equation}
\beta(t) = \frac{\sum_{i=1}^{N}n_i(t+1)-\sum_{i=1}^{N}n_i(t)}{\sum_{i=1}^{N}n_i(t)}  . 
	\label{eq: beta2}
\end{equation}
This means that $\beta(t)$ is not only the relative change of the star number
in each annulus, but also the relative change of total stars in the disc.
Replacing numerators in Equation \ref{eq: derive equilibrium equation3} with the right sides of  Equation \ref{eq: derive equilibrium equation}, we obtain analytic equations that a quasi-stationary distribution satisfies:
\begin{equation}
\begin{aligned}
    \frac{ -pn_1(t)+qn_2(t)}{n_1(t)} &= \frac{pn_1(t)-(p+q)n_2(t)+qn_3(t)}{n_2(t)} \\
    &= \frac{pn_2(t)-(p+q)n_3(t)+qn_4(t)}{n_3(t)} \\
    ... \\
    &= \frac{ pn_{N-2}(t)-(p+q)n_{N-1}(t)+qn_{N}(t)}{n_{N-1}(t)} \\
&= \frac{ pn_{N-1}(t)-(p+q)n_N(t)}{n_N(t)} .
	\label{eq: derive equilibrium equation4}
\end{aligned}
\end{equation}
To simplify Equation~\ref{eq: derive equilibrium equation4}, $n_{i}(t)/n_{i+1}(t)$ is denoted by $f_i$, where  
$i$ = 1, 2, 3, ..., $N-1$.
Adding $(p+q)$ to each fraction in Equation~\ref{eq: derive equilibrium equation4} and then dividing it by $q$, the equation becomes
\begin{equation}
\begin{aligned}
   1  + \frac{1}{f_1} &= \frac{p}{q} f_1 + \frac{1}{f_2}\\
    &= \frac{p}{q} f_2 + \frac{1}{f_3} \\
    ... \\
    &= \frac{p}{q} f_{N-2} + \frac{1}{f_{N-1}} \\
&= \frac{p}{q} f_{N-1} .
	\label{eq: derive equilibrium equation5}
\end{aligned}
\end{equation}
The equation above is a set of $N-1$ equations. When $p/q$ is given, the $f_i$ are $N-1$ unknowns and can be solved  for. After obtaining $f_i$, we can easily construct 
a quasi-stationary distribution. From this analysis, we see that the quasi-stationary distribution of the evolution is determined uniquely by the 
$p/q$ ratio, regardless of the initial distribution of stars.

When a quasi-stationary distribution is reached, the relative change of the 
star number during one timestep can be written as 
\begin{equation}
\beta(t) =\frac{-pn_1(t)+qn_2(t)}{n_1(t)}= -p +\frac{q}{f_1} .
	\label{eq: beta time}
\end{equation}
 In this limit $f_1$, and thus, $\beta(t)$ become time independent. Thenceforth, the number of stars decays 
exponentially with time  at each radius.

\subsection{Example stellar profile evolutions}

\subsubsection{Stellar profile evolution when \texorpdfstring{$p = q = 0.1$}{Lg} }
\label{sec:g stellar no bias}

\begin{figure*}
	\includegraphics[width=\textwidth]{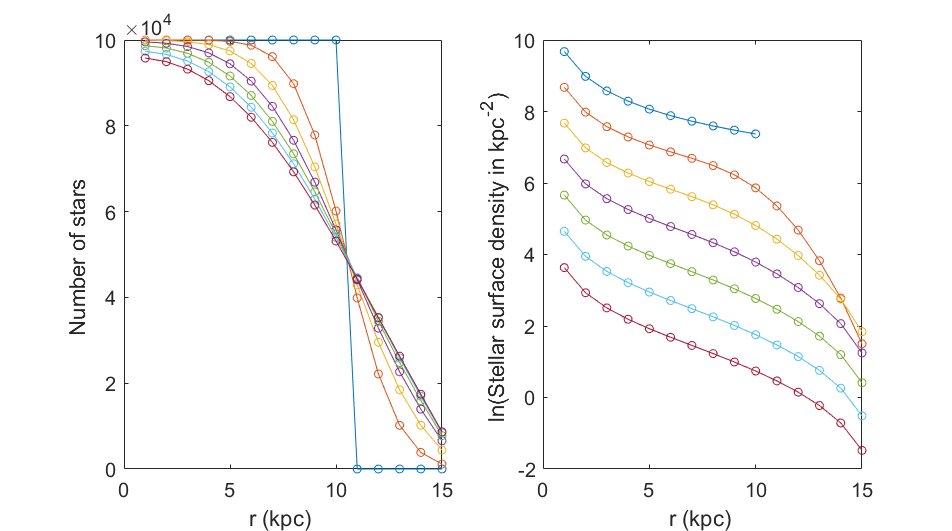}
    \caption{Time evolution of the stellar distribution from $t$ = 0
    to $t$ = 120 when $p$ = $q$ =0.1. The left panel shows the number of stars in each
    annulus and the right panel shows the surface density, i.e. the number of stars divided by the area of each annulus. On both panels, the curves from top to bottom corresponds to 
    $t$ = 0 (blue), 20 (orange), 40 (yellow), 60 (purple), 80 (green), 100 (light blue), and 120 (dark red) respectively. On the right panel, curves, except the blue one, are shifted downwards by one unit successively to give a clear view.
    }
    \label{fig:g t120 nobias}
\end{figure*}

Figure~\ref{fig:g t120 nobias} shows the number of stars and the surface density of the number in each annulus from $t$ = 0 to $t$ = 120. Due to the scattering of stars to adjacent annuli, annuli beyond 10 kpc gradually
get filled with stars and annuli within 10 kpc lose stars. As the number of stars changes in each radial bin, the surface density profile evolves with time. On the right panel, one can see that the rate of the profile change
slows down as the difference between two neighbouring curves narrows. At $t$ = 120, the surface density looks like an exponential with a cusp at the central plus a dip near the edge.

\begin{figure*}
	\includegraphics[width=\textwidth]{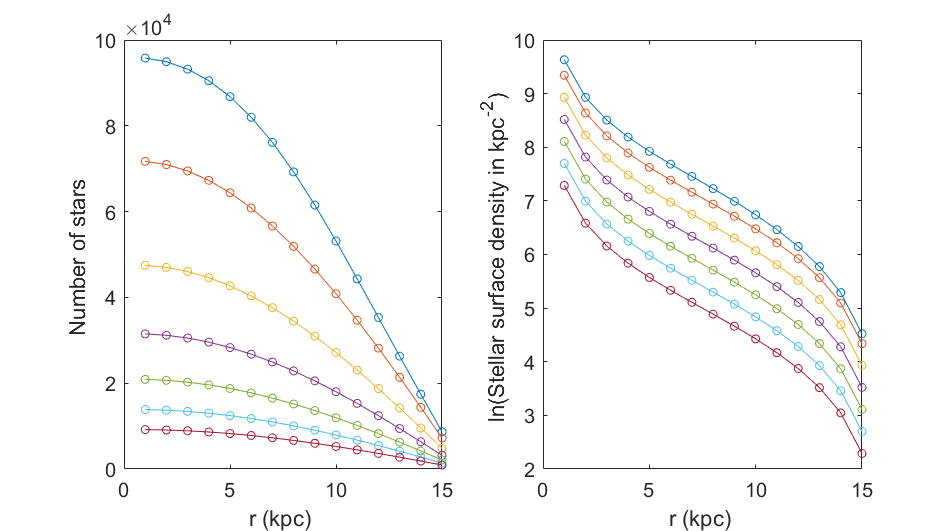}
    \caption{ Stellar profile evolution continued from Figure~\ref{fig:g t120 nobias}. 
    On both panels, the curves from top to bottom corresponds to 
    $t$ = 120 (blue), 400 (orange), 800 (yellow), 1200 (purple), 1600 (green), 2000 (light blue), and 2400 (dark red) respectively. 
    Curves on the right are not shifted
downwards artificially. The downwards motion is caused by star loss at the edge of the disc.
    }
    \label{fig:g t2400 nobias}
\end{figure*}

Figure~\ref{fig:g t2400 nobias} shows the profile evolution in a longer term from $t$ = 120 
to $t$ = 2400. Both panels show that 
the number of stars in every radial bin decreases with time.
On the right panel, the shape of the curves hardly changes, indicating
that the ratio $n_1(t) : n_2(t) : ... : n_{15}(t)$ has converged.
This means the profile has reached a quasi-stationary state at $t$ = 2400. From the right panel, one can also see that the distance 
between two neighbouring curves is almost constant after $t$ = 400. 
Since the $y$-axis is in a logarithm scale, this reveals that 
the number of stars  at each radius decays exponentially with time, as predicted  
by Equation~\ref{eq: beta time}  in the quasi-static limit.  

\begin{figure}
	\includegraphics[width=\columnwidth]{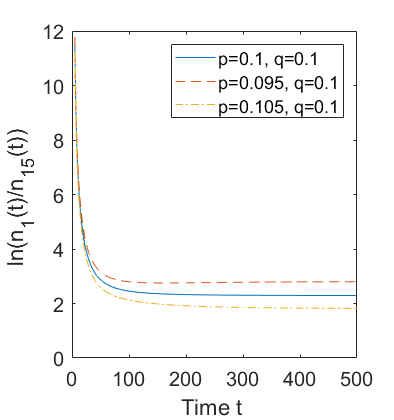}
    \caption{Time evolution of the star number ratio of the first annulus to the fifteenth annulus. The model starts
    from an $1/r$ disc with an initial size of 10 kpc. The blue solid curve, the orange dashed curve, and the yellow dash dotted curve represent different choices of $p$, $q$.
    }
    \label{fig:g nn ratio}
\end{figure}

To better understand how fast the profile converges to the
quasi-stationary distribution, we use $n_1(t)/n_{15}(t)$ as an indicator. The blue solid curve on Figure~\ref{fig:g nn ratio} shows ln($n_1(t)/n_{15}(t)$) as a function of 
time for $p = q = 0.1$. The $n_1(t)/n_{15}(t)$ ratio decreases rapidly before $t$ = 120, but the decline is hardly noticeable afterwards. This agrees with the 
profile change rate reflected by Figure~\ref{fig:g t120 nobias} and~\ref{fig:g t2400 nobias}. As the profile
approaches the quasi-stationary distribution, the $n_1(t)/n_{15}(t)$ 
ratio converges to a constant. The values of the ratio are 11.05, 9.90, and 9.87, at $t$ = 120, 500, and 1000, respectively. Beyond $t$ = 1000, the value change is less than 0.01.

\begin{figure}
	\includegraphics[width=\columnwidth]{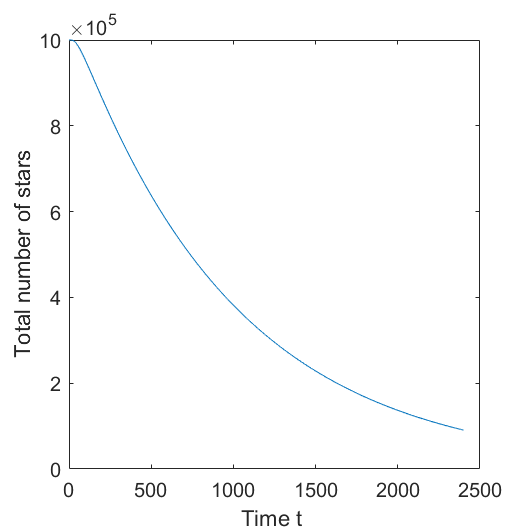}
    \caption{Time evolution of the total number of stars within 15 kpc when $p$ = $q$ = 0.1. The model starts
    from an $1/r$ disc with an initial size of 10 kpc.
    }
    \label{fig:g total star num}
\end{figure}

Figure~\ref{fig:g total star num} shows the total number of stars in the galactic disc from
$t$ = 0 to $t$ = 2400. As we mentioned before, the star loss is due to the outwards scattering at 15 kpc. When time is greater than 100, the curve confirms a clear exponential decay of stars. 
After the quasi-stationary state is reached, the relative change per timestep $\beta(t)$ is -0.0010. This value is small because
only the scattering of the outermost annulus contributes to it and 
stars in that annulus  are only 1\% of the total disc stars.
Please note that the star loss in a short term is not significant. From $t$ = 0 to $t$ = 120, the disc loses 6.47\% of initial stars.

When using the model to explain the formation of an exponential disc, we can stop 
the model at $t$ = 120 to avoid excess star losses.
A key point revealed by the model is that the scattering of stars into adjacent
annuli can lead to rapid profile changes towards a near-exponential
with the cost of losing a few percent of  the total stars. In a real
galaxy, many processes can be responsible for the scattering of stars.
If the process has a short lifetime, the star loss at the model disc edge
will stop as the process ends, so the disc will not experience the
long-term star loss.
If the scattering process does have
a long lifetime, such as scattering by spiral arms at corotation, then the long-term
star loss  rate may be exceeded by input processes, such as star formation fed by gas accretion. The losses would not be an issue in such cases either.
 We run our models for a long time in order to show detailed properties of quasi-stationary distribution rather than to present
star losses.

 The curves in Figure~\ref{fig:g t2400 nobias} can be viewed as numerical solutions to Equation
~\ref{eq: derive equilibrium equation4} or Equation~\ref{eq: derive equilibrium equation5}. In Equation~\ref{eq: derive equilibrium equation5}, the expressions on each 
 side of the equal signs all have the same form, except the first one and the last one. If the two boundary expressions got removed, 
a trivial solution would be that all the $f_i$ share the same value. 
Since $f_i$ is the star number ratio between neighbouring annuli,
this means that the star number would decay exponentially with radius and the star surface density would have the form of $e^{-r/h}/r$.
When taking the boundary expressions into account, specific restrictions are directly applied to annlui close to 1 kpc or 15 kpc.
Annlui in the middle, however, are less affected. From 5 to 10 kpc,
the star surface density should be approximately $e^{-r/h}/r$.
As $1/r$ changes much more slowly than $e^{-r/h}$, the $e^{-r/h}/r$ 
profile looks very similar to $e^{-r/h}$ in this radius range.
When looking at the right panel of Figure~\ref{fig:g t2400 nobias}, one can see that 
the surface density in this range is indeed close to a straight line,
indicating a profile of $e^{-r/h}$.

Now let's examine annuli near 1 kpc. Plugging the values of $p$
and $q$ into Equation~\ref{eq: beta time}, the equation can be written as 
\begin{equation}
 \frac{\beta(t)}{0.1} =\frac{-n_1(t)+n_2(t)}{n_1(t)}.
	\label{eq: annuli 1kpc}
\end{equation}
Since $|\beta(t)|$ is much less than 0.1, the relative change
from $n_1(t)$ to $n_2(t)$ is small, so $f_1$ is close to 1.
Using Equation~\ref{eq: derive equilibrium equation5}, we know that an $f_1$ value around 1 makes
$f_2$ close to 1 as well. It is easy to see from here that the star numbers
in annuli near 1 kpc are about the same, which gives a
surface density profile of $1/r$. This corresponds to the central
cusps on the right panel of Figure~\ref{fig:g t2400 nobias}. In a real galaxy, the  predicted overdensity at the center is usually treated as a bulge  with a distinct origin, whereas it could be the remnant of scattering in the disc.

As for annuli near 15 kpc, the last two lines in Equation~\ref{eq: derive equilibrium equation5}
indicate that $f_{N-1}$ is greater than $f_{N-2}$. If taking this
as a condition to study the relation between $f_{N-2}$ and  $f_{N-3}$,
one can obtain that $f_{N-2}$ is greater than $f_{N-3}$.
Therefore, the radial decay of the star number accelerates near
the  model disc edge, showing as the dip at 15 kpc on the right panel of Figure~\ref{fig:g t2400 nobias}. 
When combining the profile here with the part from 5 to 10 kpc, 
it resembles a Type II exponential  (see \citet{Freeman1970ApJ...160..811F}). The quasi-stationary distribution near the  model disc edge
is sensitive to the scattering in the $N$-th annulus. If the
star loss rule is modified in the model, the profile near the edge will change. For real galaxies,  stars in the outer part will continue to scatter as long as there are scattering perturbations such as gas clumps; there  may not be an absorbing boundary. Determining the model disc edge and measuring the surface density there are usually hard owing to low brightness.

\begin{figure*}
	\includegraphics[width=\textwidth]{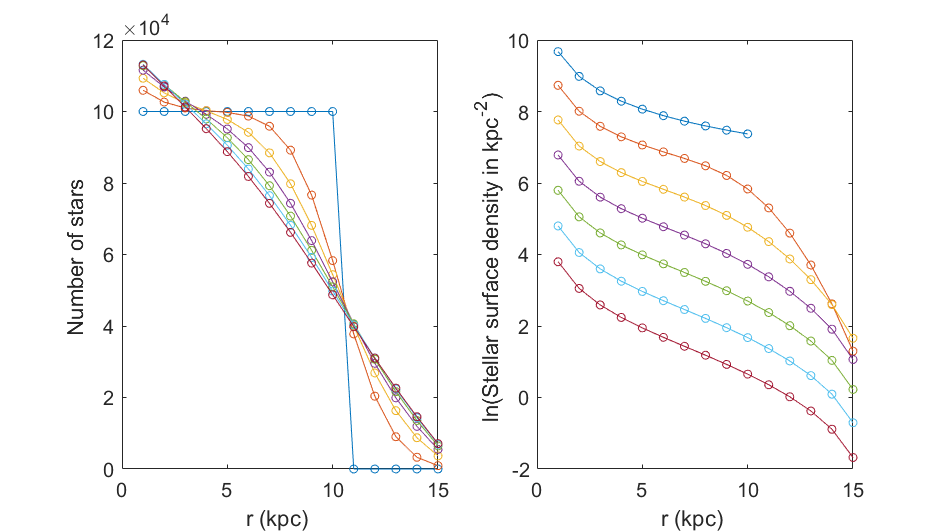}
    \caption{Time evolution of the stellar distribution from $t$ = 0
    to $t$ = 120 when $p$ = 0.095 and $q$ = 0.1. Colors of the curves represent different time in the same way as Figure~\ref{fig:g t120 nobias}. On the right panel, curves, except the blue one, are shifted downwards by one unit successively to give a clear view.
    }
    \label{fig:g t120 inward}
\end{figure*}

\begin{figure*}
	\includegraphics[width=\textwidth]{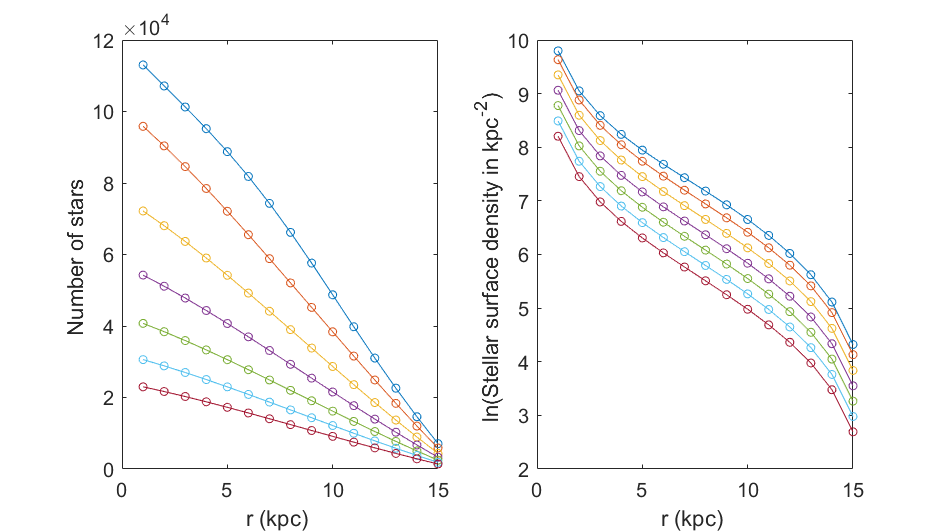}
    \caption{ Stellar profile evolution from $t$ = 120
    to $t$ = 2400 when $p$ = 0.095 and $q$ = 0.1. It is a continuation of Figure~\ref{fig:g t120 inward}. 
   Colors of the curves represent different time in the same way as Figure~\ref{fig:g t2400 nobias}. 
    Curves on the right are not shifted
downwards artificially. The downwards motion is caused by star loss at the edge of the disc.
    }
    \label{fig:g t2400 inward}
\end{figure*}

\subsubsection{Stellar profile evolution when \texorpdfstring{$p = 0.095$}{Lg} 
and  \texorpdfstring{$q = 0.1$}{Lg}}
  \label{sec:g stellar in bias}

Figures~\ref{fig:g t120 inward} and~\ref{fig:g t2400 inward} show the stellar profile changes when $p$=0.095 and $q$= 0.1.
Figure~\ref{fig:g t120 inward} illustrates a short-term evolution from $t$ = 0 to $t$ = 120, and Figure~\ref{fig:g t2400 inward} illustrates a long-term evolution $t$ = 120 to $t$ = 2400. Curves, except for $t$ = 0, on the right
panel of Figure~\ref{fig:g t120 inward} are shifted downwards by one unit successively to give a clear view.
From the two figures, one can see that the rate of the profile
evolution slows down with time and that the profile changes beyond $t$ = 120  are small.
At $t$ = 2400, the stellar number distribution has reached a
quasi-stationary state, and the corresponding surface 
density distribution is a near-exponential with a shape similar to the case of $p = q = 0.1$. From Figure~\ref{fig:g t2400 inward}, one
can also see the decay of the star number with time in every annulus due to the outwards scattering at 15 kpc.
At the quasi-stationary state, the relative star change
per timestep  $\beta(t)$ is -0.00071. Its magnitude is less than that  for  $p = q = 0.1$. The total star loss in the
short-term from  $t$ = 0 to  $t$ = 120 is 5.08\%, also less than that for  $p = q = 0.1$. These results make sense because the outward scattering at 15 kpc becomes weaker as $p$ goes from 0.1 to 0.095.

When comparing the left panel of Figure~\ref{fig:g t120 inward} to
that of Figure~\ref{fig:g t120 nobias}, a noticeable difference is that 
annuli near 1 kpc in Figure~\ref{fig:g t120 inward} gain stars while those
in Figure~\ref{fig:g t120 nobias} do not. The gain is caused by the inwards scattering bias due to $p$ < $q$. The bias also leads to
a steeper stellar gradient near 1 kpc for the
quasi-stationary state. At the quasi-stationary state, Equation~\ref{eq: beta time}
gives 
\begin{equation}
f_1 =\frac{q}{\beta(t)+p}.
	\label{eq: annuli 1kpc 2}
\end{equation}
Since $|\beta(t)|$ is very small compared to $p$, 
$f_1$ is close to $q/p = 1.053$ when $p = 0.095$ and $q = 0.1$. This means that star numbers
 in the second annulus are about 5\% less than   those in the first annulus. 
Using the first line of Equation~\ref{eq: derive equilibrium equation5}, it is easy to see that $f_1$ is close to $f_2$ when we approximate the $f_1$ on the right side using
$q/p$. Hence, a value of $f_1$ greater than 1 makes $f_2$ greater than 1 as well, so  the number of stars in the third annulus is less than that in the second annulus. 
When comparing the left panel of Figure~\ref{fig:g t2400 inward} to
that of Figure~\ref{fig:g t2400 nobias}, 
one can see that star numbers near 1 kpc decline clearly with $r$ for $p = 0.095$ and $q = 0.1$, but vary much less for $p = q = 0.1$.

The red dashed line in Figure~\ref{fig:g nn ratio} shows the time evolution of
ln($n_1(t)/n_{15}(t)$) for $p$=0.095 and $q$= 0.1. Similar to the case of $p = q = 0.1$, the $n_1(t)/n_{15}(t)$ ratio changes quickly before
$t$ = 120, confirming the relatively fast profile change in the short-term. When the quasi-stationary state is reached, the ratio converges to 16.51, which is greater than that of $p = q = 0.1$. 
This larger ratio is not only due to the steeper gradient 
in the first few annuli, as explained in the previous paragraph, but also owing to the sharper star number decrease in the middle annuli.
On the right panel of Figure~\ref{fig:g t2400 inward}, the linear fit of the quasi-stationary surface density from 6 kpc to 10 kpc gives a exponential profile with a scale length of 3.81 kpc. The fitting in the same radial interval on the right panel of Figure~\ref{fig:g t2400 nobias} gives a scale
length of 4.41 kpc. Hence, the case of $p$=0.095 and $q$= 0.1
has a steeper gradient than $p = q = 0.1$ in the middle annuli, in addition to the annuli near 1 kpc.
We will discuss the relation between $p/q$ and the scale length later
in this paper.

\begin{figure*}
	\includegraphics[width=\textwidth]{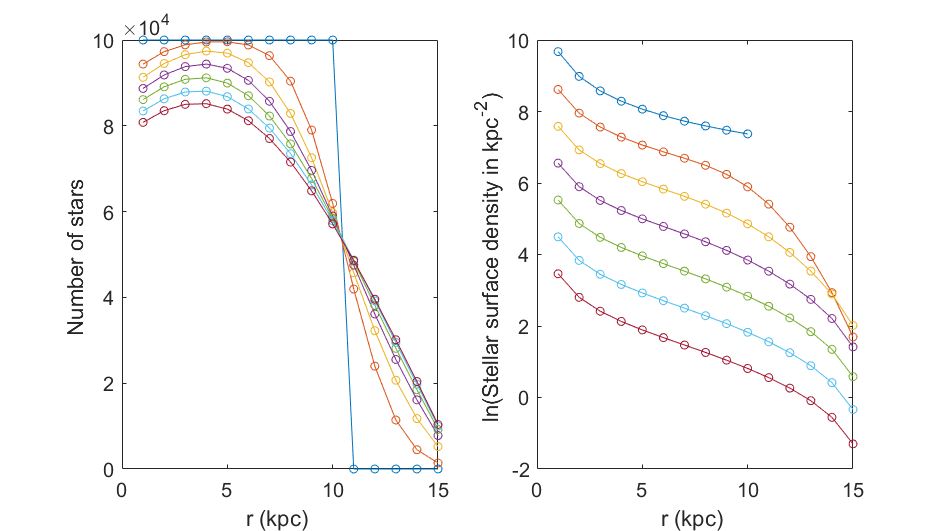}
    \caption{Time evolution of the stellar distribution from $t$ = 0
    to $t$ = 120 when $p$ = 0.105 and $q$ = 0.1. Colors of the curves represent different time in the same way as Figure~\ref{fig:g t120 nobias}. On the right panel, curves, except the blue one, are shifted downwards by one unit successively to give a clear view.
    }
    \label{fig:g t120 outward}
\end{figure*}

\begin{figure*}
	\includegraphics[width=\textwidth]{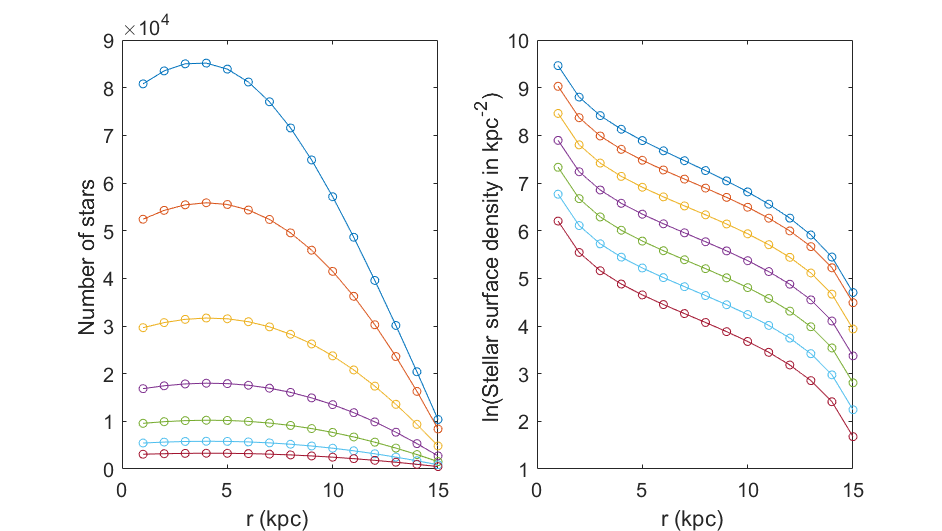}
    \caption{Stellar profile evolution from $t$ = 120
    to $t$ = 2400 when $p$ = 0.105 and $q$ = 0.1. It is a continuation of Figure~\ref{fig:g t120 outward}. 
   Colors of the curves represent different time in the same way as Figure~\ref{fig:g t2400 nobias}. 
    Curves on the right are not shifted
downwards artificially. The downwards motion is caused by star loss at the edge of the disc.
    }
    \label{fig:g t2400 outward}
\end{figure*}

\subsubsection{Stellar profile evolution when \texorpdfstring{$p = 0.105$}{Lg} 
and  \texorpdfstring{$q = 0.1$}{Lg}}
\label{sec:g stellar out bias}

After presenting the $p = q$ case and the $p < q$ case, we now illustrate a $p > q$ case. When comparing it to the $p = q$ case, many results are the opposite of the comparison between $p < q$ and $p = q$.

Figure~\ref{fig:g t120 outward} and~\ref{fig:g t2400 outward} show the profile changes for $p = 0.105$ and $q = 0.1$ from $t$ = 0 to $t$ = 120 and  $t$ = 120 to $t$ = 2400, respectively.
As before, the rate of profile changes declines with time, and
the disc converges to a quasi-stationary distribution with a
near-exponential profile. Since the outwards scattering at 15 kpc is 
stronger than the previous two cases, the star loss in the case is the largest among the three scenarios. From $t$ = 0 to $t$ = 120, the disc loses 8.08\% of the total star.
At the quasi-stationary state, the relative change of the star number during a timestep is $\beta(t)$ = -0.0014.

For $p = 0.105$, $q = 0.1$, and $\beta(t)$ = -0.0014, Equation~\ref{eq: annuli 1kpc 2} gives a $f_1$ value of 0.965. As shown in the previous subsection,
$f_1$ is close to $f_2$ at the quasi-stationary state, so $f_2$ is also less than 1. Therefore, the star numbers in the first few annuli
increase with radius as shown on the left panel of Figure~\ref{fig:g t2400 outward}.
This also means that the surface density profile near 1 kpc 
declines more slowly than the $p = 0.1$ case and the $p = 0.095$
case.

The yellow dash dotted curve in Figure~\ref{fig:g nn ratio} shows the change of $n_1(t)/n_{15}(t)$ with time for $p = 0.105$. As before, fast changes occur before $t$ = 120. The $n_1(t)/n_{15}(t)$ ratio converges to
6.1313, less than that of $p = 0.1$. The linear fit of the 
quasi-stationary surface density from 6 to 10 kpc on the right panel of Figure~\ref{fig:g t2400 outward} gives a scale length of 5.17 kpc, corresponding to a shallower gradient than $p = 0.1$.

In summary, any of the three $(p,q)$ pairs shown in this section,
regardless the direction of the radial scattering bias,
is able to generate a stellar profile almost identical to the quasi-stationary profile  within  120 time steps.  
During this time period, the star loss at the model disc edge is less than 10 \% for all the three cases. 
 The near-exponential shape of the surface density profile appears within 40 time steps. 
As the model only requires stellar scattering in each annulus, these results indicate that  many mechanisms that spread stars along the radial direction have the potential to contribute to the formation of an exponential disc. 
 The actual time length of 40 steps depends on the efficiency of the stellar scattering mechanism, i.e. how long it takes the process to scatter 10\% of stars into adjacent annuli. If the stellar scattering is due to giant clouds, Paper I indicates that 40 steps can be about 8 Gyr. If the initial stellar number distribution is centrally concentrated instead of being uniform, the time it takes to reach a near-exponential profile should be significantly less. 

\subsection{Effects from initial stellar distribution }
\label{sec:g effects}

\begin{figure*}
	\includegraphics[width=\textwidth]{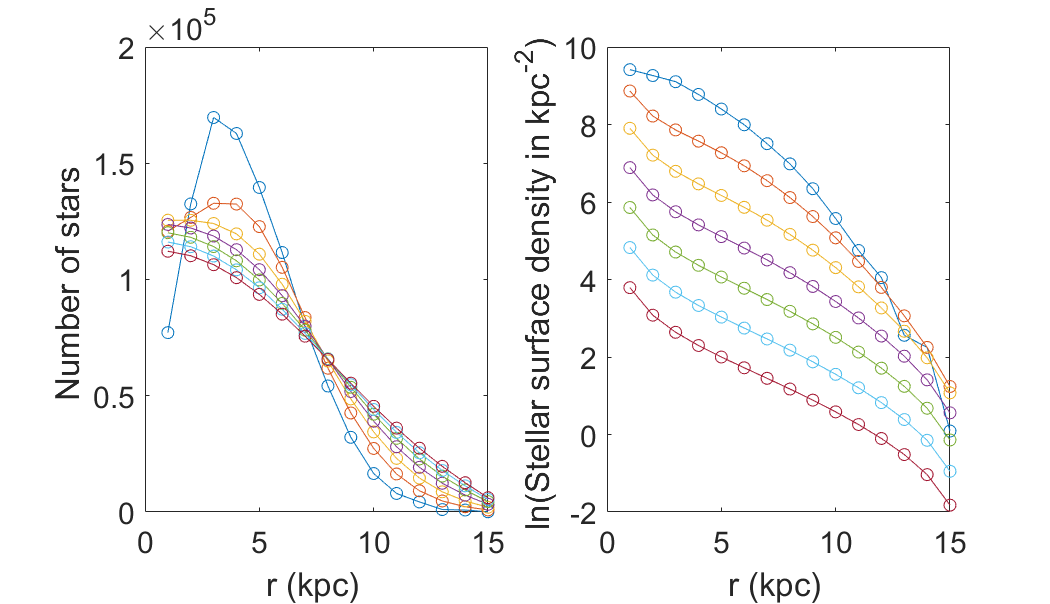}
    \caption{Stellar profile evolution of an initial S\'ersic disc with the  index $n$ of 0.5 from $t$ = 0
    to $t$ = 120 when $p$ = 0.1 and $q$ = 0.1.  A drop in stellar surface density at 14 kpc, indicating a ring-shaped pattern in the light distribution, is superposed on the initial profile. Colors of the curves represent different time in the same way as Figure~\ref{fig:g t120 nobias}. On the right panel, curves, except the blue one, are shifted downwards by one unit successively to give a clear view.
    }
    \label{fig:g 212}
\end{figure*}

\begin{figure*}
	\includegraphics[width=\textwidth]{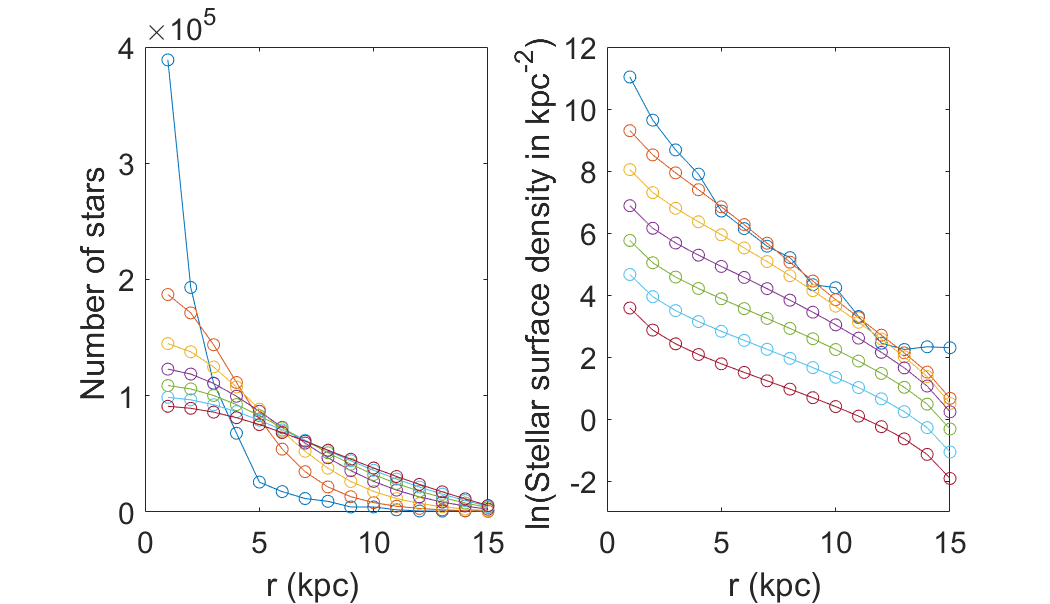}
    \caption{Stellar profile evolution of a S\'ersic disc with the index $n$ of 2 from $t$ = 0
    to $t$ = 240 when $p$ = 0.1 and $q$ = 0.1.  A few fluctuations in stellar surface density, indicating   ring-shaped patterns in the light distribution, are superposed on the initial profile. Colors of the curves represent different time. On both panels, blue, orange, yellow, purple, green, light blue, and dark red correspond to 
    $t$ = 0, 40, 80, 120, 160, 200, and 240, respectively.
    On the right panel, curves, except the blue one, are shifted downwards by one unit successively to give a clear view.
    }
    \label{fig:g t240 265}
\end{figure*}

In Section~\ref{sec:g quasi distrib}, we showed that the quasi-stationary distribution that
the model converges to is independent of the initial profile 
of stars. Nevertheless, the initial distribution has an impact on
the time needed to reach the quasi-stationary state. Results from the previous
 subsection showed the profile evolution starting from a $1/r$ disc. In this
subsection, we present profile changes with other initial profiles.

Figure~\ref{fig:g 212} shows profile changes starting from a  S\'ersic
profile with the S\'ersic index $n$ of 0.5 combined with  a drop in stellar surface density at 14 kpc. 
To be more specific, the initial S\'ersic
profile satisfies $\Sigma(r) \propto e^{-r^2/h^2}$, with a scale length $h$
of 5.4 kpc.  The drop indicates a ring-shaped pattern in the light distribution at 14 kpc. The scattering probabilities used in the model are $p = q =0.1$. Curves
in the figure reflect profiles from $t$ = 0 to $t$ = 120.
In Figure~\ref{fig:g 212}, the stellar profile evolves towards an near-exponential as expected.
At $t$ = 120, the $n_1(t)/n_{15}(t)$ ratio is 18.56, greater than 11.05, which is
the ratio at the same time in the case of $1/r$ disc.  
As the quasi-stationary ratio is 9.87, the evolution stage reached at $t$ = 120 in this trial is behind that in the $1/r$ trial. The total star loss from $t$ = 0 to $t$ = 120
is 3.86 \%.

Figure~\ref{fig:g t240 265} shows disc evolution starting from another S\'ersic
profile superposed with  a few fluctuations in stellar surface density. The S\'ersic density distribution satisfies  $\Sigma(r) \propto  e^{- \sqrt{r/h}}$, with $h$ = 0.14 kpc.
Hence, the S\'ersic index $n$ is 2. In the trial, we continue to use $p = q =0.1$.
Curves in the figure show profile changes from $t$ = 0 to $t$ = 240.
On the right panel of Figure~\ref{fig:g t240 265}, one can see that  the fluctuations in stellar surface density quickly get smoothed out and that the magnitude of the slope from 5 to 10 kpc gradually declines with time,
indicating an increasing scale length.
The $n_1(t)/n_{15}(t)$ ratios at $t$ = 120 and $t$ = 240 are 51.53 and 16.28,
respectively. Therefore, the evolution stage reached at $t$ = 240 is still behind
the $t$ = 120 stage in the $1/r$ trial. The star loss from $t$ = 0 to $t$ = 240
is 7.40 \%, which is greater than the total star loss at $t$ = 120 in the $1/r$ trial.
It is clear that this run will suffer more star loss than the $1/r$ trial when they reach the same
evolution stage. This indicates that the initial stellar distribution affects the amount of star loss.

\begin{figure}
	\includegraphics[width=\columnwidth]{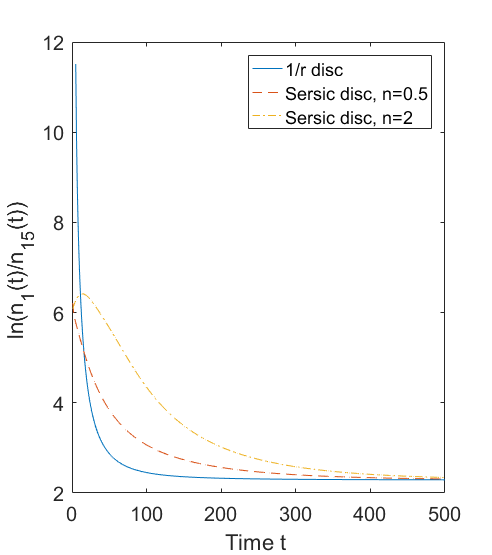}
    \caption{Time evolution of $n_1(t)/n_{15}(t)$ for different initial stellar distributions, when $p$ = $q$ =0.1.
     The blue solid curve, the orange dashed curve, and the yellow dash dotted curve correspond to initial distributions
    shown on Figure~\ref{fig:g t120 nobias}, \ref{fig:g 212}, and \ref{fig:g t240 265}, respectively. 
    }
    \label{fig:g nn ratio2 diff initial}
\end{figure}

Figure~\ref{fig:g nn ratio2 diff initial} shows ln($n_1(t)/n_{15}(t)$) as a function of time for the two runs 
presented in Figure~\ref{fig:g 212} and~\ref{fig:g t240 265}. The curve for the $1/r$ run is also added to make
the comparison easier. 
In the figure,  $n_1(t)/n_{15}(t)$ for the three runs converges  to the same value,
confirming that they share the same quasi-stationary state. Regarding the time
it takes to reach the quasi-stationary state, the $1/r$ run is the shortest and the
 S\'ersic run with $n$ = 2 is the longest. When looking at the $t$ = 0 curve
on the left panel of Figure~\ref{fig:g t120 nobias}, \ref{fig:g 212}, and \ref{fig:g t240 265}, one can see that an initial distribution 
close to the quasi-stationary distribution, i.e. the $1/r$ disc, uses a short time to converge and an initial distribution far from it, i.e. the S\'ersic profile with $n$ = 2, takes a long time to reach it. 
 Even if the two  S\'ersic runs take a longer time to reach the quasi-stationary states, the profiles start to look like a near-exponential within 40 time steps, which is comparable to the $1/r$ run.

\section{Discussion}
\label{sec:g discussion}

\subsection{Profile change rate}
\label{sec:g pro change rate}

As mentioned in Section~\ref{sec:g quasi distrib}, the quasi-stationary distribution
depends on the ratio of $p/q$. Given a fixed $p/q$, the 
magnitude of $p$ or $q$ determines the profile change rate.
Roughly speaking, the profile change rate is linearly 
correlated with the magnitude of $p$ or $q$, when $p$ and $q$ are small. The star number changes during one time step in case of $p = q =0.1$ are similar to the changes during 
two time steps in case of $p = q =0.05$.

\begin{figure}
	\includegraphics[width=\columnwidth]{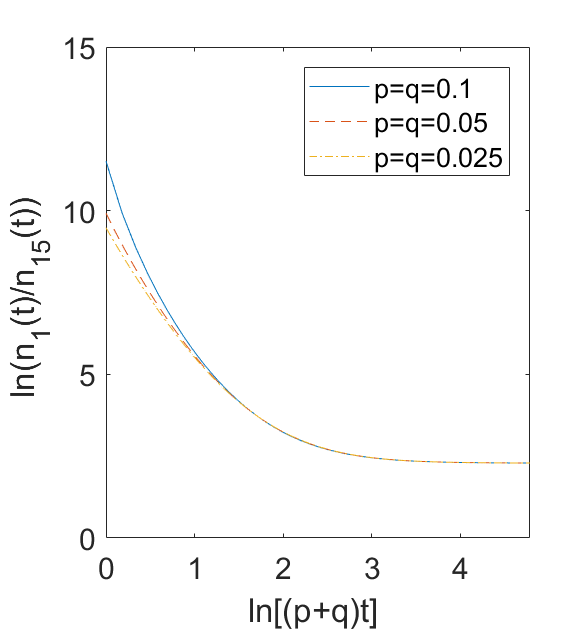}
    \caption{ Time evolution of  $n_1(t)/n_{15}(t)$ starting from the $1/r$  disc for different magnitudes of $p$ and $q$.
     The blue solid curve, the orange dashed curve, and the yellow dash dotted curve correspond to $p$ =$q$ = 0.1, $p$ =$q$ = 0.05, and $p$ =$q$ = 0.025, respectively.  ln[$(p+q)t$] is used for the horizontal axis, where the time $t$ is in unit of the number of timesteps.
    }
    \label{fig:g nn ratio3 change rate}
\end{figure}

Figure~\ref{fig:g nn ratio3 change rate} shows the time evolution of ln($n_1(t)/n_{15}(t)$)
for three runs using the same $p/q$ ratio but different $p$ values. The initial stellar profile for all three runs is 
the $1/r$ disc. 
To better see the linear relation  between the profile change rate and the magnitude of $p$ or $q$,  ln[$(p+q)t$] is plotted on the horizontal axis, instead of $t$. 
From the figure, one can see that $n_1(t)/n_{15}(t)$ of all the runs quickly converges to the same 
curve. This means that if using a certain value of $n_1(t)/n_{15}(t)$ as a milestone of the profile evolution, the time steps needed to reach the milestone in the runs are inversely proportional to $p+q$. In other words, the profile change rates are  proportional to $p+q$.

The total star losses from the beginning to ln($n_1(t)/n_{15}(t)$) = 2.40 for $p = 0.1$, $p = 0.05$, and $p = 0.025$ in Figure~\ref{fig:g nn ratio3 change rate} are 6.47\%, 6.56\%, and 6.58\%, respectively. It seems that the star loss between two evolution stages is not sensitive to $p+q$, or the profile change rate.

\subsection{Stellar profile at the disc center}
\label{sec:g disc center}

In Section~\ref{sec:g results}, we showed that the stellar surface density profile of a quasi-stationary state has an upturn  down to the center. In this subsection, we briefly discuss the shape of the upturn and its dependence on $p/q$.

\begin{figure}
	\includegraphics[width=\columnwidth]{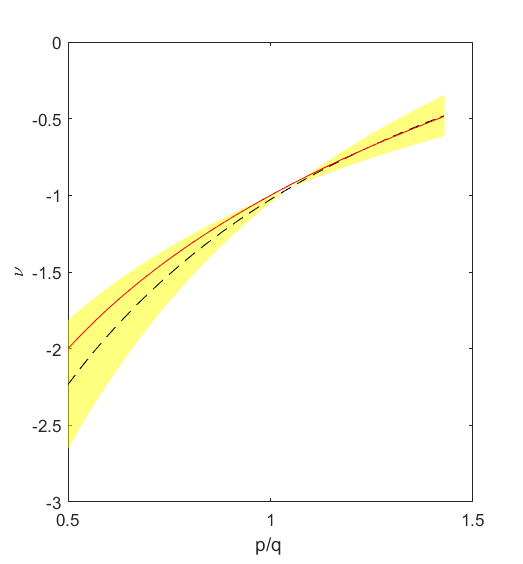}
    \caption{The power law fitting result of the quasi-stationary surface density profile at the disc center. The best fitting exponent $\nu$  is a function of $p/q$. The black dashed curve shows the best fitting exponent $\nu$ using surface density profile at $r <$ 3 kpc.  The yellow region indicates the uncertainty in the fitting results.  The uncertainty comes from two parts: the standard error of regression coefficients and the model error of using surface density profile at $r <$ 3 kpc. The latter is estimated by slightly altering the radial interval of the fitting. 
    The  red solid curve shows the values of $\nu$ estimated by Equation~\ref{eq: cent 3}.
    }
    \label{fig:nu}
\end{figure}

For simplicity, we fit the surface density $\Sigma(r)$ at low $r$ with a power law equation, i.e. 
\begin{equation}
\Sigma(r) \propto  r^{\nu}.
	\label{eq: cent 1}
\end{equation}
 As the shape of the quasi-stationary distribution depends on $p/q$, the best fitting exponent $\nu$ is a function of $p/q$. The black dashed curve in Figure~\ref{fig:nu} shows the fitting result by using the surface density profile at $r <$ 3 kpc. As $p/q$ increases, $\nu$ becomes less negative. When $p/q$ = 1, , $\nu$ is about -1. This is consistent with our analysis in Section~\ref{sec:g stellar no bias}.

We can also use Equation~\ref{eq: annuli 1kpc 2} to estimate  $\nu$. Because $|\beta(t)|$ is very small, 
$f_1$ is close to $q/p$. Using the definition of $f_1$, we obtain $n_1/n_2 \approx q/p$. As $r\Sigma(r)$ is proportional to the number of stars in the annulus at radius $r$,  the ratio of $n_1/n_2$ can be written as 
\begin{equation}
\frac{n_1}{n_2} =\frac{\Sigma(1)}{2\Sigma(2)}.
	\label{eq: cent 2}
\end{equation}
Using Equation~\ref{eq: cent 1}, $\Sigma(1)/\Sigma(2)$ is equal to 0.5$^{\nu}$. As  $n_1/n_2 \approx q/p$, Equation~\ref{eq: cent 2} becomes   $q/p \approx  0.5^{\nu}/2$, which can be rewritten as 
\begin{equation}
\nu \approx \frac{-\text{ln}(p/q)}{\text{ln}(0.5)} - 1.
	\label{eq: cent 3}
\end{equation}
The  red solid curve in Figure~\ref{fig:nu} shows the value of $\nu$ by using Equation~\ref{eq: cent 3}. It matches up with the black curve  when taking uncertainty into consideration.

\subsection{Disc scale length}
\label{sec:g disc scale len}

In Section~\ref{sec:g results}, we showed the quasi-stationary distributions for $p/q$ = 0.95, 1, and 1.05. The exponential scale lengths of these distributions measured from 6 kpc to 10 kpc are
3.81 kpc , 4.41 kpc , and 5.17 kpc respectively. In this subsection, we look at the scale length in a 
wider range of $p/q$, and study other factors in the model that can influence the scale length.

\begin{figure}
	\includegraphics[width=\columnwidth]{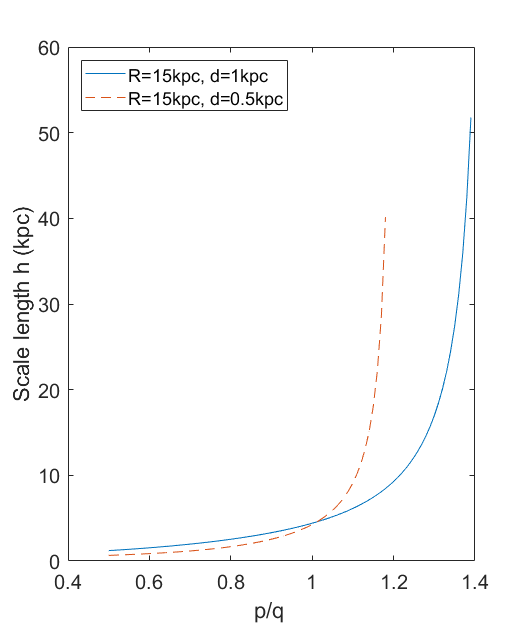}
    \caption{The quasi-stationary scale length as a function of $p/q$ for $R$ = 15 kpc.
    The solid blue curve and the dashed orange curve correspond to the annulus width $d$ = 1 kpc 
    and 0.5 kpc, respectively. The exponential scale lengths are measured from 6 kpc to 10 kpc using  linear fits.
    }
    \label{fig:g scale length r15}
\end{figure}

\begin{figure}
	\includegraphics[width=\columnwidth]{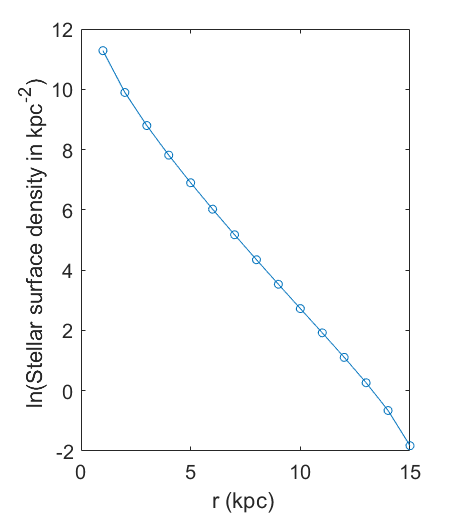}
    \caption{The  quasi-stationary stellar surface density for $p/q$ = 0.5, when $R$ = 15 kpc and $d$ = 1 kpc. 
    }
    \label{fig:g one big smooth profile}
\end{figure}

The solid blue in Figure~\ref{fig:g scale length r15} shows the scale length of the quasi-stationary state from 6 kpc to 10 kpc as a function of $p/q$, given that other parameters in the model are chosen according to the description in Section~\ref{sec:g model}.
As $p/q$ increases from 0.5 to 1.4, the scale length goes up with a growing rate, indicating that  
the surface density profile has a much shallower decline.  When $p/q$ goes beyond 1.44, the slope of the  quasi-stationary surface density from 6 kpc to 10 kpc becomes positive, leading to a density profile
increasing with the galactic radius. In contrast, when $p/q$ is much smaller than 1, 
the surface density gradient from 6 kpc to 10 kpc is close to the gradient near 1 kpc and 15 kpc, making the cusp at 1 kpc and the dip at 15 kpc hard to notice. For example, Figure~\ref{fig:g one big smooth profile} shows the  quasi-stationary surface density for $p/q$ = 0.5. The overall profile resembles
a straight line. 

 To illustrate the connection between  $p/q$ and the scale length $h$, Equation~\ref{eq: annuli 1kpc 2} is a good starting point. $f_1$, the ratio of stars in the first annulus to those in the second, will go down as $p/q$ increases, given that $|\beta(t)|$ in general is small. Ratios 
between other adjacent annuli behave in a similar way as all the $f_i$ are related by Equation~\ref{eq: derive equilibrium equation5}. Informally speaking, when the ratios decrease towards 1, the quasi-stationary surface profile becomes flatter. Thus the disk scale length $h$ rises.  By the way, when $p/q$ is greater than 1.2, the outwards scattering bias can cause a significant star loss
before the profile converges to the quasi-stationary state. When the model starts with the
$1/r$ disc, 
$p/q$ = 1.2 results in a star loss of about 45\%. On the contrary, a $p/q$ much smaller than 1 gives a star loss less than 4\%.

The disc scale length of the quasi-stationary state depends on the annulus width $d$.
The blue curve in Figure~\ref{fig:g scale length r15} uses a Markov chain model with a disc radius $R$ of 15 kpc and
an annulus width $d$ of 1 kpc. If keeping $R$ unchanged but modifying $d$ to 0.5 kpc, the number of annuli $N$ changes from 15 to 30. In this scenario,  the quasi-stationary state 
for a given $p/q$ is different, and the relation between the scale length $h$ and the $p/q$
ratio is illustrated by the dashed orange curve in Figure~\ref{fig:g scale length r15}.
As $p/q$ increases, the curve surges as the $d$ =1 kpc case but with a faster rate.  
The scale length for $d$ = 0.5 kpc is less than  $d$ = 1 kpc when $p/q$ < 1, but surpasses  $d$ = 1 kpc when $p/q$ > 1.

To understand the difference between the two curves, one can consider the star numbers at  
two galactic radii, e.g., 7 kpc and 8 kpc. When $d$ = 1kpc, these two radii correspond to
two neighbouring annulus. In contrast, when $d$ = 0.5 kpc, an  additional annulus at 7.5 kpc is placed
between 7 kpc and 8 kpc. In this case, the 
star number balance between 7 kpc and 8 kpc is maintained through the additional annulus.  The star number ratio between 7 kpc and 8 kpc is a product of the 7 kpc to 7.5 kpc ratio and the 7.5 kpc to 8 kpc ratio. When a radial scattering bias exists in the disc, this additional
annulus enlarges the star number gap  between 7 kpc and 8 kpc in the quasi-stationary distribution, as if the scattering bias doubles. 
In Figure~\ref{fig:g scale length r15}, the scale length for $d$ = 0.5 kpc  models when $p/q$ = 0.9 is 2.55 kpc, similar to the value of $h$ = 2.54 kpc for $d$ = 1 kpc  models when $p/q$ = 0.8. Likewise, 
the scale length for $d$ = 0.5 kpc when $p/q$ = 1.1 is 9.11 kpc, similar to the value of $h$ = 9.28 kpc for $d$ = 1 kpc when $p/q$ = 1.2. 
One can see that the scale length for $d$ = 0.5 kpc at a given $p/q$ is close to that for $d$ = 1 kpc at a 
$p/q$ value  about two times farther away from $p/q = 1.0$.

As mentioned in the introduction, \citet{Elmegreen2016ApJ...830..115E} studied the dependence of the scale length 
on the scattering probabilities  given an inwards bias. The scale length $0.5 \lambda/(q-p) $  given in their model increases  at a growing rate as the inwards scattering bias gets weaker, i.e. as $q$ becomes slightly greater than $p$. This qualitatively agrees with our Figure~\ref{fig:g scale length r15}, regardless of what $d$ is chosen in our model.  One difference is that their scale length 
goes to  infinity at $p/q$ =1, while our scale length does that at a
$p/q$ ratio greater than 1.
 However, this difference fades away when $d$ in our model goes to 0, as a very small $d$ makes the scale length go to infinite at a
$p/q$ ratio very close to 1. The derivation of the formula $0.5 \lambda/(q-p) $ in their model also assumes a small $d$.

The disc scale lengths are also affected by the  model disc edge $R$.
Default models use an $R$ of 15 kpc and an annulus width $d$ of 1 kpc. If we want to analyze a
larger galactic disc with an $R$ of 30 kpc, there are two basic ways to deal with it. On one hand, we can assume that the mechanism responsible for the stellar scattering scales with the  model disc edge. As increasing $R$ to 30 kpc, we also
enlarge the annulus width $d$ to 2 kpc, leaving the number of annlus $N$ unchanged.
On the other hand, one can assume that the stellar scattering is caused by a process unaffected by the global properties of the disc. In this case, we keep
the annulus width $d$ as 1 kpc, but increases $N$ to 30.

\begin{figure}
	\includegraphics[width=\columnwidth]{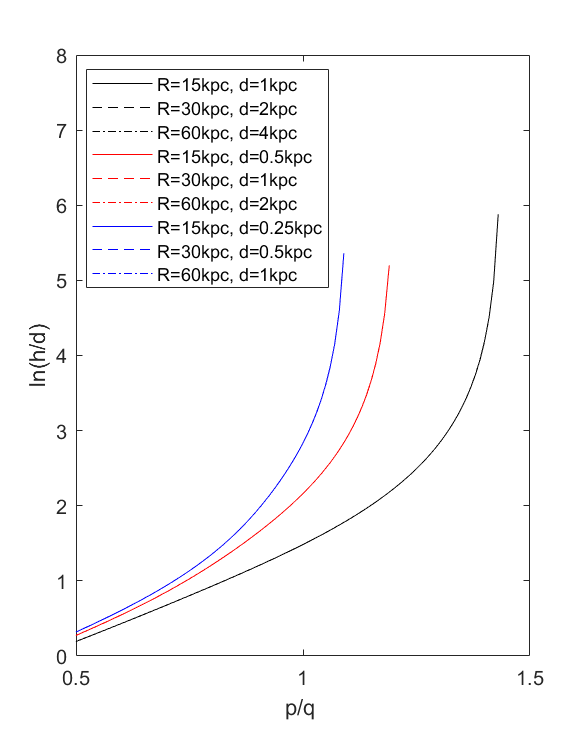}
    \caption{ The quasi-stationary scale length $h$ as a function of $p/q$ for $R$ = 15, 30, 60 kpc with various choices of $d$. Curves with the same color share the same $R/d$ ratio. To better present the curves, ln($h/d$) is plotted on the vertical axis, instead of $h$. In this way, curves with the same color overlap. 
    Nine curves are reduced to three curves, as shown in the figure.
    For $R$ = 30 and 60 kpc, the exponential scale lengths are measured from 13 kpc to 17 kpc and from 26 kpc to 34 kpc, respectively.
    }
    \label{fig:g r30 scale length}
\end{figure}

Figure~\ref{fig:g r30 scale length} shows the scale length $h$ as a function of $p/q$ for $R$ = 15, 30, 60 kpc with various choices of $d$. Curves with the same  $R/d$, i.e. the number of annlus $N$, overlap perfectly.  This indicates that as $d$ scales with 
the  model disc edge $R$, the scale length $h$ also scales with the model disc edge $R$.
When keeping the annulus width $d$ as 1 kpc, increasing $R$ from 15 kpc to 30 kpc, and then to 60 kpc, makes the  scale length $h$ go up at any given $p/q$,  i.e., as the curves go from black to red and then to blue. In this case,  the scale length changes are highly
sensitive to the value of $p/q$. For example,  from $R$ = 15 kpc to $R$ = 30 kpc,  the scale length increases at 
$p/q$ = 0.5, 0.75, 1, and 1.1 are 8.46\%, 25.1\%, 97.0 \%, and 208.8\%, respectively . 
As $p/q$ goes up, the increase surges quickly. Particularly, at $p/q$ = 1, doubling $R$ gives  a scale length increase of about 100 \%. Even if $d$ does not scale with 
the  model disc edge $R$, the scale length $h$ will roughly scale with the  model disc edge $R$ as long as $ p/q \approx$ 1.

The observation of real galactic discs shows that the exponential scale length of
spiral galaxies increases with the total stellar mass in the disc \citep{Jaffe2018MNRAS.476.4753J,Demers2019MNRAS.489.2216D}. 
Particularly, Figure 4 of \citet{Demers2019MNRAS.489.2216D} shows that the disc scale length roughly doubles when the stellar
mass goes from $10^{10.0} M_{\sun}$ to $10^{10.9} M_{\sun}$. If we assume that 
the stellar mass is proportional to the cube of the  model disc edge, a mass change from $10^{10.0} M_{\sun}$ to $10^{10.9} M_{\sun}$ corresponds to a twofold increase in the
 model disc edge. Hence, the observed disc scale length approximately scales with the 
 model disc edge. This agrees with the results from our model if the stellar orbital
shifts caused by the scattering are  treated as proportional to the  model disc edge, or if 
the shifts  are unrelated to the model disc edge but the radial scattering has little bias, i.e. $ p/q \approx$ 1.

\subsection{Time-dependent scattering probabilities}
\label{sec:g time-dep}

So far, we have treated $p$ and $q$ as two constants in the Markov chain model.  As galaxy discs evolve and consume their gas, large clumps, stochastic spiral waves and other scattering centers generally become less prevalent or smaller. Thus, the assumption of constant scattering for gigayears is not realistic in most cases. There are many possible pathways for the evolution of the populations of scattering centers.
In this subsection, we  consider some simple scenarios when $p$  and $q$ decline with time,  to illustrate some consequences when scattering in the disc weakens. 

When $p$ and $q$ are functions of time, 
we replace $p$, $q$, and the matrix $P$  in Equation~\ref{eq:trans matrix}  with
$p(t)$, $q(t)$ and $P(t)$, and Equation~\ref{eq: matrix evolution} becomes
\begin{equation}
\pi(t+1)=\pi(t)P(t).
	\label{eq: matrix evolution2222}
\end{equation}
 Here we assume that $p(t)$  and $q(t)$ decay exponentially with time and share the same decay rate.
With this choice, the $p(t)/q(t)$ ratio is independent of time and the quasi-stationary
distribution is determined by $p(0)$ and $q(0)$.  In these discrete models we achieve the exponential decay by setting $p(t)$ and $q(t)$ 
as follows:
\begin{equation}
\begin{cases}
     p(t) = p(0) \times \gamma^t, \\
    q(t) = q(0) \times \gamma^t,
   \end{cases}
	\label{eq:time depend equation}
\end{equation}
where $\gamma$ is a constant that controls the decay rate.

\begin{figure*}
	\includegraphics[width=\textwidth]{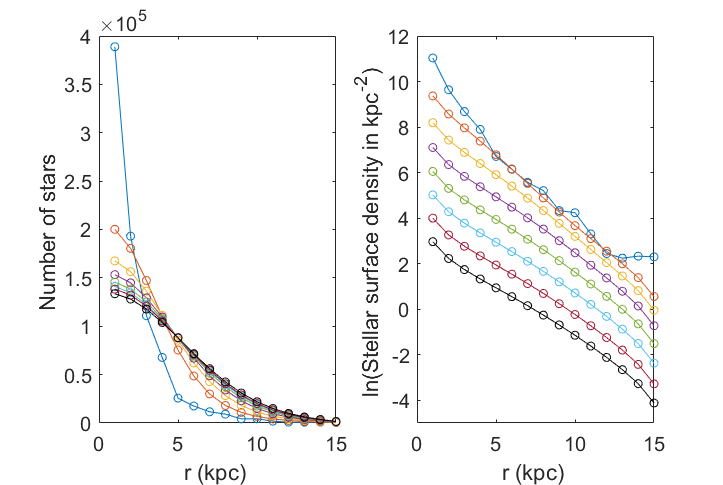}
    \caption{An example of stellar profile evolution when $p(t)$ and $q(t)$ decay exponentially with time. The initial stellar distribution in this figure is the same as that in Figure~\ref{fig:g t240 265}.
    Colors of the curves represent different time in the same way as Figure~\ref{fig:g t240 265}, except the black color, which corresponds to $t$ = 500. On the right panel, curves, except the blue one, are shifted downwards by one unit successively to give a clear view.
    Parameters used in this example are $p(0)$ = $q(0)$ = 0.1, $\gamma$ = 0.99, $R$ = 15 kpc, $d$ = 1 kpc.
    }
    \label{fig:g 265 profile with decay}
\end{figure*}

To study the effect of the time decay of $p(t)$ and $q(t)$, we start the model using the 
 S\'ersic
profile of the index $n$ = 2, which has been shown in Section~\ref{sec:g effects}.  
With the choice of  $p(0)$ = 0.1$, q(0)$ =0.1 and $\gamma$ =0.99, the stellar profile
evolution from $t$ = 0  to $t$ = 500 is illustrated in Figure~\ref{fig:g 265 profile with decay}. In the figure, the 
profile difference between $t$ = 240 and $t$ = 500 is very small, indicating that
the profile evolution ceases at $t$ = 500. When comparing this figure with Figure~\ref{fig:g t240 265},
the changes of the star number at 1 kpc show that the time-dependent model evolves more
slowly from $t$ = 0  to $t$ = 240 and that the profile reached at $t$ = 500 is close to 
that at $t$ = 120 in the time-independent model. This suggests that the $\gamma$ =0.99 model cannot converge to the quasi-stationary distribution corresponding to $p/q$ = 1.

\begin{figure}
	\includegraphics[width=\columnwidth]{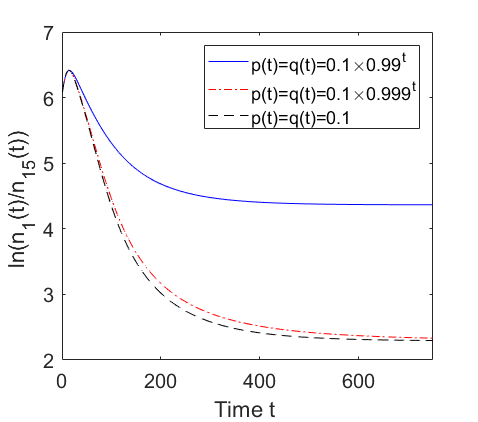}
    \caption{ Time evolution of $n_1(t)/n_{15}(t)$ in models with different $\gamma$ values.
    The solid blue curve, the dot dashed red curve, and the dashed black curve correspond to $\gamma$ = 0.99, 0.999, and 1 
 , respectively. Other parameters and the initial stellar distribution in these models are the same
 as the example in Figure~\ref{fig:g 265 profile with decay}.
    }
    \label{fig:g nn ratio with decay}
\end{figure}

To better illustrate the profile changes with time, 
 the solid blue curve and the black dashed curve in Figure~\ref{fig:g nn ratio with decay} show $n_1(t)/n_{15}(t)$ from $t$ = 0 to $t$ = 750 for the $\gamma$ =0.99 model and the  
 time-independent model, respectively.
 $n_1(t)/n_{15}(t)$ in the $\gamma$ =0.99 model converges to 78.62, much greater than
the quasi-stationary ratio of 9.87. After $t$ = 240 in the $\gamma$ =0.99 model
, $p(t)$ and $q(t)$ are less than 0.01 due to the exponential time decay.  Such small scattering probabilities are  unable to make conspicuous changes in  the stellar distribution. As a result,  the profile evolution gradually stops at an intermediate state on the way toward the quasi-stationary
state.  From the right panel of Figure~\ref{fig:g t240 265}, one can see that the profile evolution is accompanied by the increase in disk scale length. The final distribution reached by the $\gamma$ =0.99 model gives an exponential
scale length of 2.37 kpc measured from 6 kpc to 10 kpc. This is much smaller than the 
quasi-stationary scale length of 4.41 kpc,  consistent with the claim that the final distribution is an intermediate state on the way toward the quasi-stationary state.

If $\gamma$ =0.999, i.e. a slower decay than $\gamma$ =0.99, the
time evolution of  $n_1(t)/n_{15}(t)$ is presented by the red dot dashed curve in Figure~\ref{fig:g nn ratio with decay}. $n_1(t)/n_{15}(t)$ at $t$ = 750 is very close to that of the time-independent model. In this case, the disc is able to evolve to the quasi-stationary distribution before $p$ and $q$ become too small to modify the profile.

In summary, when $p(t)$ and $q(t)$ decay exponentially with the same decay rate,
the profile still evolves towards the quasi-stationary distribution, but may freeze
at an intermediate stage before reaching it. How far the profile can go on the way to the quasi-stationary distribution depends on the decay rate.  Thus, in galaxy discs the ratio of the scattering time to the timescale for the evolution of the scattering centers is an important parameter, e.g., in determining the scale length, and how much of the disc reaches a quasi-steady profile.

\subsection{Radial-dependent scattering probabilities}
\label{sec:g radial-dep}

  In the real world, scattering centers will not generally be distributed completely uniformly, so in this subsection, we briefly consider some examples where $p$ and $q$ are radially dependent. To simplify the problem, we assume here that local scattering does not have radial bias, i.e. $p_i = q_i$ for all the $i$.
We also assume that $p$ and $q$ do not vary with time.

\begin{figure*}
	\includegraphics[width=\textwidth]{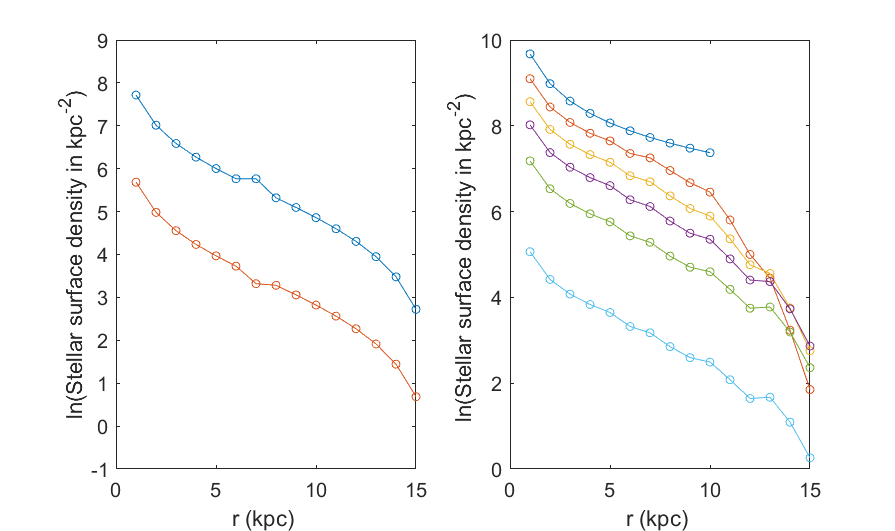}
    \caption{ Stellar profiles when $p$ and $q$ are radially dependent. The left panel shows surface density profiles where $p_i = q_i = 0.1$ except for $i = 7$.
    The top curve in this panel corresponds to 
    $t$ = 2000 (blue)  with $p_7 = q_7 = 0.08$.  The bottom curve in this panel corresponds to 
    $t$ = 2000 (orange) with $p_7 = q_7 = 0.12$. The  bottom curve is shifted downwards by two units to avoid overlapping with the top one. For both curves, the 
    initial stellar distribution is the same as that in Figure~\ref{fig:g t120 nobias}. The right panel shows an instance of surface profile evolution, if $p_i$  are randomly selected in the interval of [0.8, 1.2] and $q_i = p_i$. Here $p_i$ and $q_i$ do not vary with time. In this panel, the curves from top to  bottom correspond to
$t$ = 0 (blue), 20 (orange), 40 (yellow), 80 (purple), 400 (green), and 2000 (light blue), respectively. Curves, except the top one,
are shifted downwards by 0.5 unit successively to give a clear view. The large separation between the green curve and the light blue one  is caused by star loss at the edge of the disc.
    }
    \label{fig:radial depend p q}
\end{figure*}

We begin with the situation where scattering probabilities are the same except for one annulus.  The left panel of Figure~\ref{fig:radial depend p q}
shows $t$ = 2000 stellar profiles where the scattering probabilities at the 7th annulus are different than the rest of the disc. The quasi-stationary state is reached at $t$ = 2000, so we can treat the distributions reflected by these curves as the quasi-stationary distributions. 
 When  $q_7 = 0.08$ is less than other $q_i$  of 0.10, the lower scattering probability at the 7th annulus translates to a lower move-out rate. As a result, stars accumulate at the 7th annulus and the quasi-stationary distribution exhibits a bump there.  If  $q_7 = 0.12$ is greater than other $q_i$  of 0.10, stars at the 7th annulus are diffused and the distribution shows a pit.

The right panel of Figure~\ref{fig:radial depend p q} shows an instance if the scattering probabilities are random values selected from an interval. The resultant quasi-stationary profile has bumps and pits, so it does not follow an exponential distribution. However, if one disregards the fluctuations, it does exhibit a trend of an exponential decay with radius. These examples suffice to show that while moderate radial variations can leave an imprint, they do not disrupt the tendency of scattering to evolve toward the exponential profile.

\subsection{Scattering probabilities dependent on both time and radius}
\label{sec:g time and radial-dep}

 In this subsection, we go a little further and discuss stellar profiles where $p$ and $q$ vary with both time and radius.

\begin{figure}
	\includegraphics[width=\columnwidth]{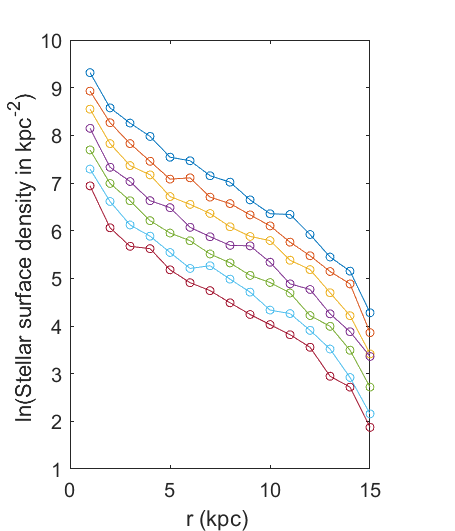}
    \caption{ 
    Stellar profile evolution of an instance where $p$ and $q$ vary with both time and radius. Here time is divided into periods of 10 timesteps. At the beginning of each period,
    $p_i$  are randomly selected in the interval of [0.8, 1.2] and  stay unchanged within the period.
    $q_i$ is chosen to be equal to $p_i$ at every radius all the time. The curves from top to  bottom correspond to
$t$ = 400 (blue), 800 (orange), 1200 (yellow), 1600 (purple), 2000 (green), and 2400 (light blue), and 2800 (dark red), respectively.    
     The 
    initial stellar distribution is the same as that in Figure~\ref{fig:g t120 nobias}. 
    }
    \label{fig:g rad and time depend}
\end{figure}

Figure~\ref{fig:g rad and time depend} shows stellar profiles where $p_i$ are randomly selected from the interval of [0.8, 1.2] every 10 timesteps and $q_i$ = $p_i$ at every radius. In Section~\ref{sec:g quasi distrib}, we show that quasi-stationary distributions are determined by values of $p$ and $q$. Here, as values of $p$ and $q$ change with time, the quasi-stationary distribution also varies every 10 timesteps.
In other words, the profile evolution will no longer converge to a single distribution.

From the previous subsection, we know that quasi-
stationary distributions of randomly selected $p_i$ 
and $q_i$ have a trend of exponential decay with 
superposed fluctuations. Curves in Figure~\ref{fig:g 
rad and time depend} show the same feature as 
expected. This confirms the robustness of the 
scattering model, i.e. 
scattering always generates an exponential trend in the stellar profile, even if scattering probabilities have moderate radial and time dependence.

\subsection{Connecting the model to specific scattering mechanisms}
\label{sec:g connecting}

As mentioned in the introduction, the stellar scattering in the radial
direction can occur in many processes of different nature. 
In this subsection, we focus on the stellar scattering caused by massive clumps in the disc, and discuss the connection between properties of clumps and  parameters of the Markov chain model. The scattering owing to other mechanisms
may be connected to the Markov chain model using a similar approach.

Key parameters used in the Markov chain model are the  model disc edge $R$, the annulus width $d$, and the scattering probabilities $p$ and $q$. 
We will discuss these parameters one by one, except $R$, which is not directly related to properties of clumps.

When the scattering is due to clumps, orbital changes of stars are caused 
by the gravitational force of clumps during star-clump encounters.
The mass of a clump determines the size of the clump neighbourhood in which the attractive force from the clump can alter stellar motions. It also controls the
amplitude of orbital shifts of incoming stars. Stellar orbits in paper I indicate
that encounters involving a clump of $7.3 \times 10^6 M_{\sun}$ can lead to a stellar radial change of about 1 kpc. With respect to this clump mass,
the annulus width $d$ can be chosen as 1 kpc so that the scattering from an annulus to adjacent annuli in the Markov chain model mimics the effect of clump-star encounters. If the clump mass is larger (or smaller) than $7.3 \times 10^6 M_{\sun}$, $d$ should be increased (or decreased) accordingly to match 
the amplitude of orbital changes.

The magnitude of scattering probabilities $p$ and $q$ in a radial bin can be connected to 
the clump number density at that galactic radius. A higher clump number density
produces more star-clump encounters during a given time period and gives
stars a higher chance to move inwards or outwards. The clump number density in the simulations of Paper I roughly corresponds to a scattering probability of 0.05 during 100 Myr.  
Paper I shows that the clump density gradient affects the scattering bias
in the radial direction. This indicates that the clump density gradient 
can be employed to determine the $p/q$ ratio. A clump profile with a steep
gradient gives an inwards scattering bias, i.e. $p/q$<1, and a shallow gradient
leads to an outwards scattering bias, i.e. $p/q$>1. In short, the clump density
and the clump density gradient at a galactic radius determine $p$ and $q$
in that annulus.

From the discussion in the previous paragraph, it is not hard to see that
scattering by clumps is location-dependent. For a given clump density profile,
the density and the density gradient of clumps usually is not the same at different galactic radius $r$. Therefore, $p$ and $q$ are usually functions of
$r$. In this paper,  we discussed only briefly the effect of a radial dependence of $p$ and $q$. This seems to be a good topic to investigate in the future. The connection between
clump profile and
the ($p$, $q$) pair also suggests that different clump density profiles can
lead to distinct profile evolution paths. When an exponential profile is  reached, the scale length $h$ may not be the same for different clump 
profiles.

Last but not least, the scattering caused by clumps can also be time-dependent. The 
{\small GADGET}-2 simulation in Paper I showed that the frequency of
clump-star encounters decline in a long term, indicating that the scattering process weakens with time. To deal with this, $p$ and $q$ in the Markov chain model can  
be written as $p(t)$ and $q(t)$ as we did in Section~\ref{sec:g time-dep}. However, the time
decay of $p(t)$ and $q(t)$ may not be exponential, so Equation~\ref{eq:time depend equation} should not be directly used. To model $p$ and $q$ as a
function of time and make it consistent with the 
{\small GADGET}-2 simulation, additional  work is needed.

In sum, the Markov chain model introduced in the paper can be treated as
a general template to study stellar profile evolution. 
It  is capable of connecting to scattering mechanisms with different nature and  
parameters in the model can be tweaked based on properties of a mechanism.

\section{Summary}
\label{sec:g summary}

The Markov chain model introduced in this paper shows that radial scattering of stars
with location-independent outwards and inwards scattering probabilities, $p$ and $q$,  leads to a quasi-stationary stellar distribution
with a near-exponential shape.
In this model, an inwards scattering bias, i.e. $p$ < $q$, is not required. 
When no bias or a slight
outwards bias is present, the stellar profile also evolves towards a near-exponential. 
This potentially allows this model to build up a stronger connection between  the exponential disc formation and many scattering mechanisms, compared with 
the stochastic models in \citet{Elmegreen2016ApJ...830..115E}, which assume an inward bias. 
Although stars in the Markov chain model can escape from the galaxy at the edge of the disc,
practical initial conditions give rise to only a small star loss, usually less than  10\% , before
the disc becomes exponential, unless a large outwards scattering bias is chosen.

The exact shape and the scale length $h$ of the quasi-stationary distribution depend on
the $p/q$ ratio, the galactic disc radius $R$, and the radial bin size $d$, but do not rely on the initial profile of stars. When $p/q$ goes from a value less than 1 to a value
greater than 1, the quasi-stationary scale length increases with a growing rate.
For fixed values of $p/q$ and $d$, the expansion in $R$ always causes an increase in $h$, but the magnitude
of the increase is related to $p/q$.  
When the disc radius $R$ doubles, the scale length $h$ will double if $p/q \approx$  1 or if the radial bin size $d$ doubles with $R$.  This scaling relation between $R$ and $h$ qualitatively agrees with
results from observations  \citep{Demers2019MNRAS.489.2216D}.
 Given $R$, the variation in quasi-stationary scale length due to a fixed change in the radial scattering bias $p/q$ becomes less  as $d$ increases.

The time it takes to converge to the  quasi-stationary  distribution depends on the initial 
stellar profile and the magnitudes of $p$ and $q$. An initial profile far from the quasi-stationary distribution or  small  magnitudes of $p$ and $q$ lead to a long time of
convergence. If the magnitudes of $p$ and $q$ decay with time, profile evolution can 
stop at an intermediate stage, failing to get to the presumed  quasi-stationary state. As the radial profile change
is very rapid at the beginning of a Markov chain run, the intermediate stage may have already reached a near-exponential shape, although different from that of the  quasi-stationary state.

Most of the results of this paper have been derived in the context of constant or near constant scattering probabilities $p, q$ across the model disc, though some spatial and temporal variations were considered in Section 4. It was shown that bumps and dips formed with spatially dependent scattering probabilities, could persist for some time. This raises the possibility that surface density profiles could be tuned to a variety of persistent forms by specific gradients in the scattering probabilities, which could result from gradients in the number or masses of the scattering centers. However, we note that the evolution of lumps and dips is inherently self-limiting, rather than self-reinforcing. E.g., in a zone with low scattering probabilities, the excess influx from adjacent zones would build a lump, but the excess density in the lump would increase the outflow from that zone until the scattering fluxes balance. Moreover,  if stellar scatterings are caused by massive scattering centers, the distribution of scattering centers is itself likely to be subject to self-regulation. For example, a region with small or few scattering centers, would increase its surface density from an unbalanced inflow, which can lead to enhanced local gravitational instability and the formation of new, massive scattering clumps. The specific timescales are important, and non-exponential profiles may get frozen in if self-regulating processes are slow, as discussed above. The complexities of these issues are beyond the scope of the present paper, but merit further study.



\section*{Data availability}

The data used in this article will be shared on request to the corresponding author.




\bibliographystyle{mnras}
\bibliography{reference} 








\bsp	
\label{lastpage}
\end{document}